\documentclass[sigconf]{acmart}

\AtBeginDocument{%
  \providecommand\BibTeX{{%
    \normalfont B\kern-0.5em{\scshape i\kern-0.25em b}\kern-0.8em\TeX}}}

\copyrightyear{2025}
\acmYear{2025}
\setcopyright{none}

\acmConference[FAccT '25]{ACM Conference on Fairness, Accountability, and Transparency}{June 23--26,
  2025}{Athens, Greece}

\usepackage{caption}
\usepackage{subcaption}
\usepackage{float}
\usepackage{enumitem}
\begin{document}

\title{The Value of Disagreement in AI Design, Evaluation, and Alignment}


\author{Sina Fazelpour}
\affiliation{%
  \institution{Northeastern University}
  \city{Boston, Massachusetts}
   \country{United States}
}
\email{s.fazel-pour@northeastern.edu}

\author{Will Fleisher}
\affiliation{%
  \institution{Georgetown University}
  \city{Washington, D.C.}
   \country{United States}
}
\email{will.fleisher@georgetown.edu}

\renewcommand{\shortauthors}{Sina Fazelpour and Will Fleisher}

\begin{abstract}
    Disagreements are widespread across the design, evaluation, and alignment pipelines of artificial intelligence (AI) systems. Yet, standard practices in AI development often obscure or eliminate disagreement, resulting in an engineered homogenization that can be epistemically and ethically harmful, particularly for marginalized groups. In this paper, we characterize this risk, and develop a normative framework to guide practical reasoning about disagreement in the AI lifecycle. Our contributions are two-fold. First, we introduce the notion of perspectival homogenization, characterizing it as a coupled ethical-epistemic risk that arises when an aspect of an AI system's development unjustifiably suppresses disagreement and diversity of perspectives. We argue that perspectival homogenization is best understood as a procedural risk, which calls for targeted interventions throughout the AI development pipeline. Second, we propose a normative framework to guide such interventions, grounded in lines of research that explain why disagreement can be epistemically beneficial, and how its benefits can be realized in practice. We apply this framework to key design questions across three stages of AI development tasks: when disagreement is epistemically valuable; whose perspectives should be included and preserved; how to structure tasks and navigate trade-offs; and how disagreement should be documented and communicated. In doing so, we challenge common assumptions in AI practice, offer a principled foundation for emerging participatory and pluralistic approaches, and identify actionable pathways for future work in AI design and governance.

\end{abstract}





\maketitle

\section{Introduction}\label{sec:intro}
Artificial intelligence (AI) systems have rapidly permeated many spheres of contemporary life. The potential of these systems to fundamentally shape people's lives---and their rights and duties---has focused research and regulatory attention on evaluating and governing the societal impacts of AI adoption~\cite{tabassi2023artificial,jobin2019global}. These efforts have significantly advanced our understanding of how central concerns such as reliability and validity~\cite{coston2023validity,jacobs2021measurement}, bias and fairness~\cite{fazelpour2021algorithmic,barocas2023fairness}, transparency and explainability~\cite{fleisher2022understanding,vredenburgh2022right}, privacy~\cite{hoofnagle2012behavioral,nissenbaum2019contextual}, and accountability~\cite{raji2020closing,lazar2024legitimacy} interact with sociotechnical AI systems. However, an important yet underexplored issue in these discussions involves the homogenization that occurs when \textit{disagreements} that arise during AI design and evaluation are unjustifiably discarded. This homogenization risks producing systems that reflect dominant viewpoints in ways that undermine both the general performance of these systems, and the interests of marginalized groups affected by them. In this paper, we conceptualize the risks of such homogenization. We then develop a normative framework to guide practical reasoning about disagreement in AI pipelines, in order to realize the epistemic value of such disagreement.

Disagreements are pervasive throughout the AI lifecycle, arising in contexts such as problem formulation~\cite{passi2019problem} and dataset construction~\cite{uma2021learning} as well as model design~\cite{davani2021dealing}, evaluation~\cite{ganguli2022red}, and alignment~\cite{bai2022training}. Yet, recent empirical analyses show how standard practices in AI design and development systematically disregard the significance of these disagreements, resulting in a homogenization that particularly harms marginalized communities~\cite{fleisig2023majority,narimanzadeh2023crowdsourcing,davani2021dealing}. Annotator disagreements in tasks such as hate speech and toxicity detection offer salient examples. Such disagreements can be rooted not only in varied interpretations and cultural norms, but also in differences in experiences, expertise, and situated knowledge~\cite{waseem2016you}. Yet, by systematically discarding divergent opinions, standard aggregation methods like majority voting not only disregard the input of those potentially most knowledgeable, but can also result in systems whose failures disproportionately harm marginalized communities~\cite{sap2019risk,dixon2018measuring}. These problems are only exacerbated by upstream selection practices that already under-sample the most pertinent perspectives~\cite{fazelpour2022diversityml}, and downstream documentation and communication practices that report only aggregated labels and no trace of disagreements~\cite{prabhakaran2021releasing}. Similar problems permeate the entire AI lifecycle. 

\begin{figure*}[ht]
     \centering
     \includegraphics[width=\linewidth]{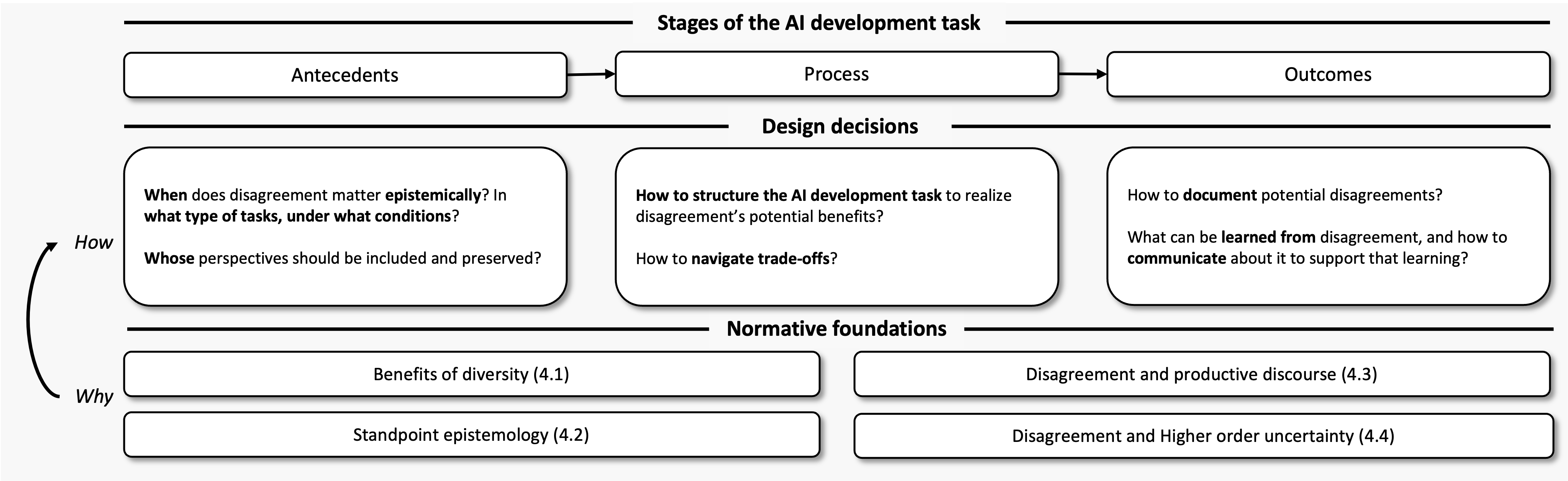}
    \caption{A schematic depiction of the proposed framework for reasoning about how disagreement at different stages of AI development tasks---such as data labeling and annotation, red-teaming, ruleset elicitation for value alignment---wherein divergences in opinions or perspective may be expected. In all such tasks, researchers and practitioners face challenging questions at different stages of tasks about potential epistemic value of disagreements, how best to proceed in a given stage to realize those values (Figure inspired by Figure 1 in \cite{huang2024collective}.)}
        \label{fig:disagreement_stages}
\end{figure*}

A nascent line of research has begun to develop alternative sociotechnical approaches that address the critical shortcomings of standard practices for handling diversity and disagreements. These approaches emphasize the need to enhance the diversity of perspectives involved in the AI lifecycle, preserve disagreements, and amplify the voices of those whose views might otherwise be discounted~\cite{basile-etal-2021-need,braun2024beg,cabitza2023toward,fleisig2024perspectivist,fazelpour2022diversityml}. While these approaches have made significant contributions, key questions remain about both their normative justification and their appropriate practical application~\cite{fleisig2024perspectivist}. 

In this paper, we propose a normatively grounded way of reasoning about and realizing the value of disagreement in AI development
pipelines, thereby supporting efforts to resist harmful homogenization. Our contributions are two-fold. First, we introduce the notion of \textit{perspectival homogenization} and characterize it as a \textit{coupled ethical-epistemic} risk that arises when an aspect of an AI system's design, evaluation, or alignment processes unjustifiably diminishes or eliminates disagreement and diversity of perspectives. We show how this conception provides a unifying lens for understanding the interplay between disagreement, on the one hand, and broader concerns about fairness, accuracy, and validity, on the other. We further argue that perspectival homogenization is best understood as a \textit{procedural} risk, which calls for targeted interventions throughout the AI development lifecycle.

Second, we develop a normative framework to guide such interventions. Specifically, drawing on research in feminist and social epistemology, philosophy of science, psychology, and organizational science, we identify four lines of research that articulate distinct \textit{epistemic} rationales for valuing disagreement---rationales grounded in the knowledge-related benefits that disagreement and diversity can provide. These rationales highlight the positive role of diverse perspectives and disagreement in fostering collective intelligence, improving informational exchange quality, expanding shared evidential resources, enhancing the quality of collective inquiry, offering higher-order evidence, and more. In each case, we clarify the mechanisms that underpin these benefits, and specify the conditions under which they are most likely to emerge.

To structure the application of these insights to AI design and governance, we distinguish between three stages in \textit{AI development tasks} where divergent views are likely to arise. These tasks include activities such as data annotation, red-teaming evaluation, and alignment rule construction. Each stage of these tasks raises key disagreement-related design decisions (see Figure~\ref{fig:disagreement_stages} for a schematic overview): 
\begin{itemize}
    \item The \textit{antecedent stage}: Can disagreement offer epistemic benefits in the task at hand, when, and under what conditions? Which perspectives should be selected to promote those benefits? 
    \item The \textit{process stage}: How should the task and communication be structured to appropriately support the process of disagreement? How should trade-offs, such as communication friction or time inefficiencies, be navigated? 
    \item The \textit{outcomes stage}: How should disagreement be documented? What can be learned from disagreement, and how can communication about disagreement support that learning? 
\end{itemize}

By linking each of these design decisions to relevant epistemic rationales for valuing disagreement, we identify critical correctives to mistaken, but widely held assumptions in AI practice. For instance, our arguments challenge the view that disagreement is relevant only in \textit{subjective} tasks and only on political or ethical grounds, showing instead how disagreement also plays a crucial epistemic role in a wide range of \textit{complex, objective tasks}. We also clarify the distinction between mere demographic membership in marginalized groups and the possession of epistemically valuable \textit{standpoints} associated with those groups, helping to guide more principled approaches to participation. Moreover, we highlight two key levers for realizing disagreement's benefits in discourse, while managing its costs: modifying the \textit{structure} of communication---through network topologies that strategically shape who is exposed to which perspectives---and modifying the \textit{content} of communication---by expanding it from surface-level judgments to the justifications that support them. More broadly, the framework developed here contributes to emerging efforts, including works in perspectivist approaches to annotation, participatory AI, and pluralistic alignment, by providing a normative foundation for these methodologies, and offering actionable pathways for future work.

\section{Related work}\label{sec:background}
Disagreement is prevalent across the AI design and governance lifecycle. In problem formulation, disagreements often arise when determining which characteristics define socially beneficial AI systems~\cite{bergman2024stela,lee2019webuildai} and when operationalizing contested or ambiguous concepts (e.g., a ``successful employee'')~\cite{kleinberg2018discrimination,passi2019problem}. In dataset construction, annotators frequently diverge in their assessments due to differences in interpretation, cultural contexts, expertise, lived experiences, and more~\cite{kapania2023hunt,sachdeva2022assessing,uma2021learning,zhang2015perceived}. For instance, in hate speech detection, there can be significant disagreements between labels by domain experts and those annotated by crowd workers~\citep{waseem2016you}. Similarly, disagreements permeate AI evaluation and alignment processes, whether through discrepancies in human-annotated benchmark datasets~\citep{jha2023seegull}, diverging preferences and expectations regarding appropriate model behavior~\cite{casper2023open}, or differing assessments among those tasked with identifying model vulnerabilities~\cite{ganguli2022red}.

Yet, these disagreements tend to be systematically obfuscated and dissolved in standard AI design and evaluation practice, leading to epistemic and ethical concerns. For example, typical methods for resolving annotator disagreement, such as majority voting and averaging, privilege majority opinion and suppress minority viewpoints. In this way, they can result in datasets that (primarily) reflect the majority or dominant viewpoints, leading to systems with reduced generalizability across different social and cultural contexts~\cite{davani2021dealing,fleisig2023majority}. These epistemic shortcomings can, in turn, produce ethical failures, such as biased and unfair outcomes that disproportionately affect marginalized communities. Examples include models that disproportionately mislabel online content in African American English as toxic or fail to categorize homophobic content as hate speech~\cite{sap2019risk,dixon2018measuring}. These issues are exacerbated by selection and entry mechanisms that tend to hinder the participation of relevant perspectives from the outset, resulting in a less diverse set of experts, annotators, or evaluators~\cite{fazelpour2022diversityml,fleisig2024perspectivist}. Furthermore, the absence of robust documentation practices to track disagreement obscures the extent of contestation within datasets used in training and evaluation~\cite{prabhakaran2021releasing}. These challenges underscore the need for a systematic approach to dealing with disagreement throughout the AI pipeline.

A growing body of work aims to address these issues by introducing sociotechnical approaches that are better capable of eliciting diverse viewpoints, and capturing and incorporating disagreement. Examples include participatory-oriented efforts to address underrepresentation~\cite{kirk2024prism}, include diverse collectives~\cite{huang2024collective} or center marginalized communities~\cite{bergman2024stela} in pluralistic alignment; machine learning methodologies such as soft labeling~\cite{peterson2019human}, jury learning~\cite{gordon2022jury}, and multi-annotator modeling~\cite{davani2021dealing}; enhanced dataset documentation frameworks that recognize and report diversity~\cite{diaz2022crowdworksheets}; and conceptual frameworks such as perspectivism, which provide resources for embracing the multiplicity of human perspectives in data labeling and model training~\cite{cabitza2023toward,fleisig2024perspectivist}. These approaches have been promising, resulting in epistemically \textit{and} ethically improved systems~\cite{davani2021dealing,narimanzadeh2023crowdsourcing,lee2019webuildai}. Nonetheless, progress has been hindered by the lack of normative frameworks that offer guidance on key questions, such as when disagreements are productive, whose perspectives should be included and preserved, and how we should leverage the benefits of disagreements given potential trade-offs~\cite{fleisher2022understanding}. In this paper, we offer such a framework along with conceptual tools to unify key concerns in this line of work, while also providing important correctives to assumptions that are widespread even in this more recent line of research.  

Before proceeding, it is important to distinguish homogenization of the sort we are concerned with here from related concepts such as outcome homogenization~\cite{bommasani2022picking}, which might result from algorithmic monoculture~\cite{kleinberg2021algorithmic}. These notions are primarily concerned with the homogenizing effects of adopting \textit{many} identical or sufficiently similar AI models across an ecosystem. In contrast, what we refer to as perspectival homogenization concerns how disagreement is handled within a \textit{single} AI system's development pipeline, potentially affecting the system even if deployed in an environment with no other AI system or with very different systems. As such, perspectival homogenization need not imply algorithmic monoculture or outcome homogenization. 

To further see the distinction, consider a set of content moderation systems, each suffering from perspectival homogenization, in the sense that, by ignoring dissenting perspectives, each exhibits failures that reflect the blindspots of a particular majority. This is compatible with these perspectivally homogeneous systems being quite different and making very different mistakes from one another (i.e., not exhibiting monoculture or outcome homogenization). This can happen, for instance, when the sampled perspectives, and so the particular majority, substantially differ across systems (e.g., drawn from populations with very distinct geographical or linguistic expertise). Of course, things would be different, if those perspectives were also the same or substantially similar. How different types of homogenization relate to one another thus depends on other upstream or downstream choices and considerations in the AI lifecycle. Future work should examine the potentially reinforcing interactions between these concerns by situating them within the broader context of discussions around diversity and pluralism in the AI pipeline~\cite{fazelpour2022diversityml,jain2024algorithmic}. 
\section{Perspectival homogenization}\label{sec:homogenization}

\subsection{Perspectival homogenization as a coupled ethical-epistemic concern}
We propose the notion of perspectival homogenization to capture the related set of challenges surrounding dealing with disagreements in AI design, evaluation, and alignment processes, as exemplified by the issues discussed above. Perspectival homogenization occurs when an \textit{aspect} of an AI system's design, evaluation, or alignment processes \textit{excludes} or \textit{attenuates} the impact of \textit{relevant} disagreement or diversity of perspectives concerning the task at hand. Whether they occur intentionally or not, perspectivally homogenized aspects of an AI development pipeline privilege dominant or majority perspectives, while filtering out or discounting dissenting or minority viewpoints. 

We use the term ``aspect'' to capture the broad range of methodological choices and assumptions that can lead to such homogenizing effects. These aspects can include technical methods for opinion aggregation, such as majority voting and averaging, but also extend to sociotechnical choices such as selecting which domain experts or populations to consult, determining who is included in dataset construction or model evaluation pools, and establishing what information to report about dataset properties, model predictions, or performance. Perspectival homogenization can result from decisions made at any of these stages, such as undersampling perspectives from certain groups, or failing to effectively elicit their views. 

These aspects may either ``exclude'' or ``attenuate'' disagreement and diversity. In data labeling, for instance, the use of a majority rule aggregation method can result in the input of annotators with a minority opinion being completely eliminated. 
However, homogenization can also occur through attenuation, where the influence of a minority viewpoint is merely diminished rather than eliminated. Consider a regression task predicting a continuous value, where annotators provide numerical inputs that are aggregated using a weighted averaging method. In such cases, the dominant perspective disproportionately influences the final outcome, potentially leading to misleading conclusions. Similar impacts result from sampling decisions that occur before aggregation, and shape the annotator pool's relevance for the task at hand.

Crucially, we emphasize that the disagreement or diversity of perspectives must be ``relevant'', as not all forms of disagreement contribute to epistemically and ethically better outcomes. Determining what constitutes relevant diversity and disagreement---and consequently, identifying unjustified instances of homogenization---requires a clearer understanding of what makes disagreement productive. We explore these considerations in the following sections.

Finally, we characterize perspectival homogenization as a \textit{coupled ethical-epistemic} concern, in the sense developed by philosopher Nancy Tuana~\cite{tuana2010leading,tuana2013embedding,tuana2017understanding}. Such concerns include cases where ``value decisions embedded in research models and methods ... go unquestioned and often unappreciated,'' affecting \textit{both} ethical and epistemic values in intertwined ways~\cite[][p. 1957]{tuana2013embedding}. Rendering such issues transparent and critically examining their coupled ethical-epistemic significance has been shown to enhance the objectivity and reliability of scientific inquiry in domains ranging from climate modeling to public health policy to diversity research~\cite{tuana2010leading,tuana2013embedding,tuana2017understanding,steel2018multiple,katikireddi2015coupled}. As discussed in the previous section, perspectival homogenization exemplifies such an issue, given its origin in often underappreciated methodological choices and assumptions throughout the AI lifecycle, leading to both epistemic and ethical downstream consequences. What is more, here too, attending to ethically motivated considerations regarding the inclusion of diverse perspectives---whether through enhancing participation of relevant perspectives~\cite{lee2019webuildai,jha2023seegull} or by going beyond majority voting when learning from those perspectives~\cite{davani2021dealing,narimanzadeh2023crowdsourcing}---has been shown to lead to functionally improved models vis-a-vis considerations of accuracy and robustness.

\subsection{Perspectival homogenization as a procedural risk}
Perspectival homogenization can thus result in AI systems that exhibit worse epistemic (e.g., accuracy) and ethical (e.g., fairness) performance. Nonetheless, we believe there are three important sets of reasons for treating perspectival homogenization as a \textit{procedural risk}---that is, a risk that should be assessed and addressed throughout AI design and evaluation processes, rather than just in terms of the problematic outcomes it may bring about in relation to the AI products of those processes. 

First, perspectival homogenization provides a form of \textit{diagnostic specificity} that enhances the quality of mitigation and management interventions. Recent work in responsible AI underscores that addressing issues related to accuracy, bias, and fairness is often best achieved by targeting potential sources of the problem throughout the AI design and development pipeline~\cite{coston2023validity,black2022algorithmic}. In contrast, mitigation strategies (e.g., adding fairness constraints in optimization) that fail to address underlying sources---such as sampling biases---may not only prove ineffective but could also adversely impact the very groups they aim to protect~\cite{fazelpour2022algorithmic,lipton2018does}. Since perspectival homogenization locates the source of certain challenges to epistemic and ethical values in how disagreement is dealt with in different AI development tasks, addressing it as it arises within the task at hand is crucial.

Second, \textit{epistemic feasibility} can necessitate a procedural perspective towards addressing perspectival homogenization. The AI pipeline consists of many distinct tasks---such as problem formulation, dataset construction, model evaluation, user study, and more---which are often carried out by different teams, possibly from distinct organizations~\cite{fazelpour2022diversityml,winecoff2024improving}. While these teams have control over how disagreement is addressed in their respective task (e.g., in dataset construction), it is not feasible to expect them to have the same access or control over how their handling of disagreements impact downstream model behavior. Treating homogenization as a procedural risk allows for targeted, task-appropriate interventions without imposing unrealistic expectations.

Third, and finally, addressing perspectival homogenization ensures the \textit{procedural integrity} of AI development processes, which is valuable in \textit{its own right}. For example, regardless of downstream outcomes, there is epistemic value in making implicit assumptions explicit, justifying them, and subjecting them to scrutiny~\cite{longino1990science}. As discussed in the next section, fostering active disagreement can play a crucial role in this regard. Furthermore, while our primary focus is an epistemic one, it is important to note that acknowledging and incorporating disagreement and diverse perspectives is also a key consideration in procedural fairness~\cite{zhang2015perceived,bohman2007political}.

Treating perspectival homogenization as a procedural risk has important implications; it underscores the need for AI risk management and auditing frameworks to include systematic evaluations for how disagreements are handled throughout AI pipelines, in turn highlighting the need for a normative framework to guide such evaluations and related interventions. Even frameworks such as NIST's AI Risk Management Framework~\cite{tabassi2023artificial} that emphasize the need for a process-oriented approach to risk management (e.g. for validity or bias) overlook the issue of disagreement across the AI lifecycle. While recent efforts for dealing with disagreement---such as multi-annotator modeling and soft-labeling techniques---can help, they currently lack a coherent normative framework to guide their implementation. Without clear principles to determine when disagreements are valuable and whose perspectives should be preserved, these methods risk being applied inconsistently, ineffectively, or worse inappropriately. By providing structured guidance to support these emerging approaches, a normative framework ensures they align with relevant ethical and epistemic goals that motivate them. We now turn to developing such a framework.

\section{The Epistemic Value of Difference and Disagreement: Why}\label{sec:why-disagree}
As discussed in Section~\ref{sec:background}, divergent views and opinions are likely to emerge in many AI development tasks, from data annotation to model evaluation to alignment. This section, together with the next, develops a normative framework to guide reasoning about disagreement in such tasks---aimed at mitigating the risk of perspectival homogenization and realizing the benefits of disagreement. To organize our discussion, we distinguish between three stages at which disagreement-related decisions arise within such AI development tasks: an \textit{antecedent stage}, where decisions about relevance and inclusion are made; a \textit{process stage}, where choices about task structures and communication dynamics shape how disagreement and potential trade-offs are handled; and an \textit{outcomes stage}, where the implications of disagreement for downstream use and communication are evaluated (see Figure~\ref{fig:disagreement_stages}). Disagreement and diversity can be valuable at each of these stages, in ways that are grounded in varied, and distinct, epistemic, moral, and political rationales.\footnote{By a rationale, we mean a set of reasons that support the promotion or realization of a specific value.} Understanding the rationales for why disagreement matters---and the explanatory mechanisms that underpin these rationales---offers critical insight into how practitioners might leverage the design choices at each stage to realize these benefits in practice.

We focus specifically on \textit{epistemic} rationales for valuing diversity and disagreement---those that highlight their knowledge-related benefits. This is not to downplay the importance of other reasons for valuing disagreement and contestation, such as their political significance~\cite{landemore2017beyond,mouffe2000deliberative,mutz2006hearing,anderson2006epistemology}. Rather, we center epistemic rationales because, despite their importance, these considerations and their significant implications are often underappreciated in AI research on disagreement. Moreover, getting clear about these rationales can help when thinking about other reasons for valuing disagreement, as even theorists who foreground the political value of disagreement frequently regard its epistemic benefits to be central to their view---e.g., highlighting disagreement's role in enhancing the quality of reasoning and justification practices~\cite{anderson2006epistemology,landemore2017beyond}.

In what follows, we review four lines of research that provide distinct epistemic rationales for disagreement and outline the underlying mechanisms that explain when and why those benefits can be expected. While some of these rationales pertain primarily to design choices at a single stage of the AI development task, others are relevant across multiple stages. We return in Section~\ref{sec:how-disagree} to the key design decisions introduced above, demonstrating in each case how the insights from the rationales relevant to it can guide more principled reasoning.

\subsection{Epistemic benefits of diverse perspectives}
\label{subsec:diversity}
Disagreement in AI development tasks often arises through interaction among individuals with diverse perspectives.\footnote{This is not to say that diversity of perspectives is either necessary or sufficient for disagreement.} To appreciate how such disagreement can be epistemically productive, it is essential to understand when and why the underlying diversity of perspectives can lead to epistemic advantages. This understanding can inform key design questions---such as when and in what types of tasks disagreement is likely to be beneficial, whose perspectives should be included to enable such benefits, and how task structure should be designed to support them. In this section, we focus on research from philosophy, psychology, organizational science, and social science that examines the epistemic benefits of diversity in social identities, broadly construed to include demographic attributes, disciplinary backgrounds, and more.\footnote{Our discussion here closely follows~\citet{fazelpour2022diversityml}. We focus on social identities in this broad sense, because they offer a useful proxy for ``perspectives'' (e.g., disciplinary backgrounds and experience might inform ways of thinking about an issue).}

Following~\citet{steel2019information}, we can distinguish two primary pathways through which diversity enables epistemic benefits: cognitive pathways and information elaboration pathways. The first, cognitive or task-related pathways, concern how identity-based diversity contributes to beneficial differences in the cognitive resources individuals bring to a task~\cite[for a comprehensive review, see][]{page2017diversity}. People with different social backgrounds are likely to possess distinct skills, areas of expertise, background knowledge, or domain-specific experiences. In group settings, these differences expand the overall pool of task-relevant resources and can lead to complementary problem-solving strategies. For example, consider annotators labeling a dataset of dialogues with a healthcare AI system. Such tasks require diverse cognitive inputs, including background knowledge, case-specific information, conceptual representations, and inferential reasoning. A more diverse group is likely to draw on a wider and more varied set of knowledge bases, represent problems differently, and apply alternative heuristics—thereby reducing correlated errors and expanding the epistemic reach of the group. Importantly, the epistemic value of diversity through these pathways depends on the nature of the task~\cite{fazelpour2022diversityml,hong2004groups}. Diversity is particularly beneficial for tasks marked by complexity, uncertainty, or the need for situated knowledge. In contrast, for routine or low-skill tasks, sociocultural differences may have little impact on the distribution of relevant cognitive skills.

The second mechanism, information elaboration pathways, concerns how groups elicit, share, examine, and integrate information distributed across their members~\cite{fazelpour2022diversityml,steel2019information}. Even when a group is cognitively diverse, its ability to realize epistemic benefits depends on how effectively members communicate and critically engage with one another's contributions towards the shared task. Research shows that sociocultural homogeneity undermines these processes in several ways~\cite{phillips2009pain,fazelpour2021diversity}. For example, individuals in socially homogeneous groups tend to assume cognitive similarity with others who share their sociocultural background. This leads to overestimation of shared knowledge and views, and failures in eliciting or sharing critical information~\cite{phillips2017real,roberson2019diversity}. In contrast, members of more socially diverse groups are more likely to expect epistemic differences, which motivates both more thorough informational exchange and better preparation in \textit{anticipation} of possible disagreement~\cite{phillips2009pain}. Moreover, social homogeneity increases conformity pressure. Individuals in such groups may suppress dissenting views to signal belonging or avoid social sanction, even when group consensus is illusory~\cite{deutsch1955study,cialdini2004social}. Homogeneous groups also show greater resistance to disagreement from in-group members~\cite{phillips2017real}. Sociocultural diversity has been shown to reduce these pressures and increase the expression of dissenting views~\cite{phillips2006surface}. 

Of course, communication in socially diverse groups is not without its costs---such as friction between perspectives, or communication barriers across disciplinary or cultural lines. The extent to which diversity yields epistemic benefits via information elaboration pathways depends crucially on factors like communication structure~\cite{fazelpour2021diversity,heydari2015efficient} and the surrounding organizational culture~\cite{phillips2017real,fazelpour2022diversityml}. In particular, egalitarian institutional culture that explicitly values diversity is needed to create environments where individuals are more likely to voice their opinions and engage critically with one another~\cite{phillips2017real,bear2011role}.

\subsection{Insights from standpoint epistemology}
\label{subsec:standpoint}
Feminist epistemology provides further evidence of the epistemic benefits of diversity and disagreement. These include the insights of feminist standpoint theorists concerning the epistemic advantages possessed by people in marginalized communities \cite{intemann201025,Toole2021-TOORWI,harding2004feminist,harding2009standpoint,pohlhaus2002knowing}.\footnote{Standpoint theory can be seen as extending the benefits of diversity discussed in the last section, focusing in particular on contexts where the relevant perspectives concern socially marginalized but epistemically advantaged standpoints.} Standpoint theory suggests that certain forms of diversity can be \textit{particularly} epistemically beneficial: specifically, diversity that results from the inclusion of people with socially marginalized but epistemically advantaged \textit{standpoints}. In certain contexts, people inhabiting these standpoints possess both greater knowledge and improved abilities to gather and evaluate evidence. In general, standpoint theorists share a commitment to three main claims \cite{Toole2021-TOORWI,intemann201025}. First, what individuals are in a position to know depends on facts about their history and social situation (the \textit{Situated Knowledge Thesis}). Second, that a standpoint is a set of epistemic resources achieved by collective consciousness raising, a process of critical reflection on, and resistance to, oppressive structures (the \textit{Achievement Thesis}).  Third, the standpoints achieved by (and concerning) marginalized groups can confer an epistemic advantage on those possessing the standpoint (the \textit{Epistemic Advantage Thesis}).

According to the situated knowledge thesis, an individual's social situation influences not just what evidence and experiences they have, but also their epistemic resources for evaluating their evidence and experiences \cite{Saint-Croix2020-SAIPAP-4}. These epistemic resources can include concepts, abilities, and strategies for gaining knowledge in a domain, and their availability depends on the needs a group of people to describe and navigate their lived social realities~\cite[p.~341]{Toole2021-TOORWI}. A standpoint consists in a set of such epistemic resources. 

The achievement thesis claims that standpoints must be achieved through active collective work of critically engaging with and resisting the forms of oppression faced by a specific marginalized group. The inclusion of the achievement thesis avoids any essentialist baggage, allowing that both group members and non-members alike can gain a particular standpoint associated with that group. This achievement requires gaining the experience of resisting the relevant kind of oppression, either explicitly or implicitly, which in turn leads to the development of the relevant conceptual and interpretive resources \cite{intemann201025,collins1986learning,Wylie2003-WYLWSM}. Crucially, despite being open to non-member standpoint attainment, the achievement thesis retains the connection between the standpoint and the social situation of the group in question. This is because it requires that a subject engage in reflection on the social circumstances of that group. Moreover, achieving a standpoint requires taking part in group inquiry or consciousness-raising with members of the relevant group \cite{Harding2004-HARASR-4}. 

Finally, the epistemic advantage thesis claims that subjects occupying a standpoint (often) have better access to knowledge, at least with respect to certain subject matters or domains of knowledge~\cite{wylie2017knowers}. A person with the right standpoint has epistemic resources, including concepts and cognitive abilities, that others, including those who are beneficiaries of systems of oppression, lack. For instance, a Black woman in the US with the concomitant standpoint will be better at recognizing certain patterns of behavior than people lacking the standpoint. She may notice that certain behavior comprises harmful microaggressions, or that certain seemingly positive characterizations (e.g., ``Strong Black Woman'') can constitute a harmful stereotype~\cite{castelin2022ma}. This is as a result of having developed epistemic resources such as concepts (e.g., hate speech, misogynoir \cite{bailey2021misogynoir}),
habits of attention, recognitional competences, and inductive methods. 

The upshot of the foregoing discussion is that we should expect members of marginalized groups to have better access to knowledge concerning features of their social circumstances. This includes knowledge about actions that help to shape oppressive social systems. This last point is of crucial importance for thinking about the dangers of perspectival homogenization. Standpoint theory suggests that homogenization is not simply a matter of losing some valuable perspectives, but of excluding epistemically critical standpoints. The epistemic damages of perspectival homogenization can thus be worse in cases where there are relevant standpoints, as in the case of hate speech annotation, or more generally application domains with a long history of institutional failures for members of a particular community~\cite{friesen2022standpoint}.\footnote{%
    While we endorse standpoint theory, note that weaker claims can support a similar kind of argument for the epistemic benefits of disagreement. Feminist empiricism, for instance, appeals to differences in evidence to explain much of the same phenomena as standpoint theorists \cite{intemann201025} Liam Kofi Bright argues that any empiricist should have many similar commitments to standpoint theory given the evidence in the actual world \cite{bright2024duboisian} See also our earlier discussion of diversity (sec. 4.1).
} 

\subsection{Disagreement and productive discourse}
\label{subsec:disagreement}
In Section~\ref{subsec:diversity}, we highlighted the epistemic benefits of expected cognitive difference and disagreement through information elaboration pathways. In this section, we focus on the \textit{activity} of disagreement: the interaction between people who hold differing opinions and engage in discussion or debate. Disagreement in discourse can provide several distinctive epistemic benefits. 

First, the process of disagreement generates expectation that disagreeing interlocutors will provide reasons or evidence for their side of the disagreement~\cite{Fleisher2019-FLEEAA-4,Fleisher2020-FLEHTE-2}. If I assert a claim and you deny it, I expect you to justify your denial---and you, in turn, expect me to justify my initial assertion. This shared norm motivates disagreeing parties to bring forward richer evidence, increasing the amount and depth of information available to the group as a whole. The result is that groups characterized by disagreement are more likely to possess a better common stock of evidence. 

Second, disagreement can promote an epistemically valuable division of cognitive labor, where disagreeing parties are motivated to search for further evidence to support their position~\cite{Mercier2011-MERAIA,Cruz2013-CRUTVO}. At the same time, they are motivated to more carefully scrutinize the opposing position. As a result, groups that disagree are better at solving logical problems like the Wason Selection Task~\cite{Mercier2011-MERAIA}. In this context, behaviors resembling confirmation bias---a well-known individual vice---can contribute to better group inquiry. Because each side is committed to defending its own position and challenging the other’s, disagreement sustains inquiry over time and helps preserve minority views that may otherwise be prematurely dismissed.

This insight echoes a broader theme in philosophy of science: that a well-structured division of cognitive labor can increase a collective's likelihood of reaching truth~\cite{Kitcher1990-KITTDO,Strevens2003-STRTRO-5,Thoma2015-THOTED-2,muldoon2013diversity,Fleisher2018-FLERE}. Feminist philosophers have extended this idea, arguing for a \textit{procedural} conception of epistemic objectivity that requires structured processes where participants with differing background beliefs and values can critically evaluate each other's positions~\cite{longino1990science,harding2019objectivity}. Disagreement in such settings may also prompt the search for additional alternative hypotheses, especially in contexts where these hypotheses help protect important beliefs or values~\cite{solomon2007social}.

While these mechanisms are similar to the information elaboration pathways discussed in Section~\ref{subsec:diversity}, they offer a distinct set of epistemic benefits: they highlight advantages that arise specifically by including people who actively disagree with one another concerning the topic at hand, and providing them with the opportunity to express and contest that disagreement. Social diversity may increase the likelihood of such disagreement, but it is neither necessary nor sufficient. To reap these epistemic gains, it is often necessary to deliberately create space for disagreement through task design or organizational practices. Again, these benefits will be lost in cases of perspectival homogenization.

\subsection{Disagreement and higher-order evidence}
\label{subsec:preservation}
A final line of research highlights the epistemic benefits of disagreement in terms of the higher-order evidence it provides. In epistemology, a distinction is commonly drawn between first-order evidence, which bears directly on beliefs about the world, and higher-order evidence, which concerns the reliability of one's evidential situation, rational capabilities, or reasoning procedures that underpin the formation of those beliefs~\cite{sep-higher-order-evidence}. Disagreement is a paradigmatic form of higher-order evidence: discovering that others---especially well-informed or similarly situated others---disagree with you about a claim is often a reason to reduce your level of confidence---an intuition that has received sustained discussion in the epistemology of disagreement~\cite{Christensen2009-CHRDAE-2,Matheson2015-MATTES-6,sep-disagreement,Fleisher2020-FLEHTE-2}.

Uncertainty estimation is an important aspect of making rational decisions. Determining when it is permissible or appropriate to deploy a predictive algorithm will depend on appropriate estimations of how accurate the system is. Keeping track of higher-order evidence concerning how confident one should be in such estimations of accuracy is itself important for making rational decisions. This is because such higher-order uncertainty should be taken into account in one's first-order confidence about the truth of the claim in question. If I become highly confident that a hate speech classification system reliably classifies certain types of social media post, only to later find out that a trusted collaborator is much less confident in that regard, this should give me reason to alter my own initial estimate of the system's reliability. Similarly, learning that a dataset involves significant disagreements about certain types of posts is higher-order evidence that, all else equal, should reduce one's confidence in the reliability of a system trained using that dataset.

Beyond the fact of disagreement, the content communicated during disagreement can also carry evidential value. The quality of the reasons offered in support of a claim during deliberation (as discussed in~\ref{subsec:disagreement}) can inform assessments of a claim's reliability. In cases where aggregating divergent views is desirable, such assessments can serve as valuable input for reliability-weighted integration methods~\cite[e.g.,][]{colombo2017bayesian}. In addition to offering insights into the quality of reasoning, disagreement can also surface important information about the variety of reasoning. For instance, if multiple individuals converge on the same conclusion using diverse lines of reasoning or a variety of distinct evidential resources, then, all else equal, that convergence may offer stronger confirmatory support compared to convergence among a less methodologically or evidentially diverse group~\cite{landes2021variety,heesen2019vindicating,kuorikoski2016evidential}. 

Reasoning about why disagreement exists---not just that it exists---can thus offer valuable higher-order insights, particularly when evaluating the robustness of a conclusion or deciding how much weight to assign to different viewpoints. Conversely, the failure to preserve or communicate details about disagreement amounts to a loss of valuable evidence, or worse a distortion of how much confidence a process and its outcomes warrant. A further way that perspectival homogenization is epistemically harmful, then, is by eliminating evidence about disagreements in decision processes, impeding downstream users' ability to calibrate their confidence. 
\section{The Epistemic Value of Difference and Disagreement: How}\label{sec:how-disagree}
In the previous section, we identified four lines of research that offer distinct rationales for \textit{why} disagreement can be epistemically beneficial. In this section, we aim to bridge the gap between those insights and their practical implications for \textit{how} to realize the benefits of disagreement in the relevant AI development tasks. We show how attending to the mechanisms underpinning these rationales can guide practitioners' reasoning about key design decisions encountered across AI development tasks, where disagreement is likely to arise (see Figure~\ref{fig:disagreement_stages}). While our recommendations are not exhaustive, they offer a feasible starting point for efforts to deal with disagreement in ways that realize its epistemic value and avoid perspectival homogenization. Throughout, we demonstrate how this framework provides a robust normative foundation for existing methodologies, while also identifying mistaken assumptions and suggesting new directions for research.

\subsection{When disagreement epistemically matters: Beyond ``subjective'' tasks}
In what kinds of AI development tasks can practitioners expect diversity and disagreement to yield epistemic benefits? As \citet{fleisig2024perspectivist} observe, a common assumption in research on disagreement in AI is that disagreement's relevance is largely confined to subjective tasks. This assumption also underlies recent work in pluralistic alignment~\cite[e.g.,][]{kirk2024prism}. The reasoning seems to be that when an objective or independent ``fact of the matter'' exists, diverse perspectives are unnecessary, and disagreements merely reflect error---noise to be averaged away. In contrast, in subjective tasks where objective criteria are taken to be absent, a plurality of values may be at play, and disagreement is treated more seriously, often as a matter of fair and inclusive participation. On this view, disagreement matters on ethical or political grounds, but not epistemic ones.

Our first recommendation, then, is to resist this assumption and to recognize that disagreement can carry epistemic significance even in tasks not typically viewed as subjective. As discussed in Section~\ref{sec:why-disagree}, the value of disagreement (and diversity) is not confined to---nor even primarily found in---contexts of subjective preference or value divergence. The four lines of research examined above all concern the \textit{epistemic} value of disagreement and diversity: they focus on improving performance in tasks such as prediction, explanation, problem-solving, and forming accurate beliefs based on appropriate reasons. As noted in Section~\ref{subsec:diversity}, the epistemic value of diverse perspectives and disagreement is particularly salient when such tasks are marked by complexity, uncertainty, or the need for situated knowledge---features that characterize a wide range of AI development tasks.

These characteristics are frequently present even in tasks conventionally classified as objective. For example, in the labeling and annotation of clinical data, there can be significant and epistemically meaningful disagreement among domain experts~\cite{valizadegan2012learning,schaekermann2019understanding,elmore2015diagnostic}. Similarly, red-teaming evaluations of high-risk systems---such as those involving biological security threats or healthcare misinformation---are often framed as objective exercises in vulnerability discovery. Yet, reliably performing these tasks requires drawing on diverse forms of expertise, including technical knowledge, contextual awareness, system-level mental models, relevant threat models (the capabilities of malicious users in security risks, or the psychology of different vulnerable user groups for misinformation risks), and experience with adversarial testing strategies~\cite{Anthropic,ahmad2025openai}. They are precisely the types of tasks where cognitive diversity and epistemically relevant disagreement can play a critical role. 

In short, the scope of tasks in which disagreement should be taken seriously must be significantly expanded. Efforts to realize the epistemic benefits of disagreement---such as those discussed below---are thus relevant far beyond subjective domains.

\subsection{Whose perspectives should be included: Standpoints \textit{vs.} demographic attributes}\label{subsec:why-perspectives}
Not all divergent viewpoints contribute to epistemically productive disagreement, nor do all minority positions merit preservation. A key design decision, then, concerns which perspectives to include in AI development tasks in order to foster the conditions under which disagreement's epistemic benefits can emerge. Attending to the mechanisms discussed in Section~\ref{sec:why-disagree} can again provide valuable guidance in navigating this selection decision. To see this, let us focus on a response to this question---common in participation-oriented AI research and pluralistic alignment---which prioritizes the inclusion of individuals from marginalized perspectives. This emphasis is well-motivated: the exclusion of marginalized perspectives often leads to epistemically harmful blind spots and unfair outcomes. Conversely, their inclusion can provide unique task-relevant insights. However, careful attention to the details of standpoint theory is essential to appropriately realize these epistemic benefits. 

As discussed in Section~\ref{subsec:standpoint}, a standpoint is not merely a reflection of one's demographic identity; it is \textit{achieved} through collective deliberation about, critical engagement with, and resistance to oppressive structures in relation to specific issues. While members of marginalized groups are more likely to attain such standpoints, they do not do so simply in virtue of demographic membership. Conflating standpoints with demographic attributes does not just incorrectly essentialize members of marginalized groups---e.g., by treating every individual from a group as if they should have expertise on causes and consequences of the particular forms of oppression facing that group across domains. It also risks missing out on the epistemic benefits that standpoints can offer, while leading to a false confidence about inclusion. This creates a serious disconnect between the normative rationale provided by standpoint theory and how it is implemented in practice.

Take the example of annotating hate speech. If disagreement among annotators is resolved via majority vote, and annotators from a socially marginalized but epistemically advantaged standpoint constitute a minority, their input---grounded in lived experience, situated knowledge, and collective engagement with relevant issues---may be systematically discounted. This results in perspectival homogenization, where the insights of those best positioned to identify harm are lost. Crucially, however, this situation cannot be corrected simply by including more annotators from that marginalized demographic, if those individuals have not had opportunities to cultivate the standpoint-relevant expertise~\cite[see also][]{waseem2016you,kapania2023hunt}. The epistemic risks here are twofold: missing relevant insights and overestimating the value of inclusion.

Recent work in participation-oriented AI has developed promising methodological infrastructure to support the inclusion of perspectives from marginalized communities in different aspects of the AI lifecycle, such as the creation of benchmark evaluation datasets~\cite{jha2023seegull} or the development of norms for aligning AI systems~\cite{bergman2024stela}. Pursuing these efforts in ways grounded in the framework proposed here can help ensure that the epistemic rationale for including standpoints is properly reflected in practice. Specifically, future work could operationalize inclusion not merely in terms of demographic representation, but in terms of building partnerships with leaders and experts from relevant communities, civil society organizations, and advocacy groups. Doing so increases the likelihood of including individuals with relevant standpoints.\footnote{Also relevant here are proposals regarding when and how such participation is most fruitful~\cite[e.g.,][]{suresh2024participation,delgado2023participatory}.}

Finally, while this subsection has focused on standpoint theory and its implications for the selection question, the discussion of cognitive benefits of diversity (in~\ref{subsec:diversity}) is also highly relevant~\cite[see also][]{fazelpour2022diversityml}. To further operationalize those diversity-based rationales, future efforts can draw on research in cognitive and team task analysis~\citep{schraagen2000cognitive,cooke2013interactive,page2017diversity}, which offers guidance for identifying task-relevant knowledge and designing teams that benefit from the complementary skills of their members. 

\subsection{How to structure the task: Independent individuals \textit{vs.} networked collectives}\label{subsec:why-task}
In many AI development tasks where disagreement arises, task participants are asked to work \textit{independently}, and in isolation from one another. Divergent opinions are often treated as unproductive disagreement or noise---an obstacle to be averaged away or disregarded altogether. Yet, as discussed in Section~\ref{sec:why-disagree}, the epistemic benefits of diversity and disagreement do not stem from aggregating isolated judgments. Rather, they emerge when individuals operate as parts of networked collectives. This requires an epistemic infrastructure that is organized to bring their diverse perspectives to bear on a shared task through interaction and deliberation, under appropriate norms.

Consider the typical setting in which participants work independently and disagreements are resolved through majority voting or averaging. This pattern is currently common for a variety of AI development tasks, including data labeling, annotation of AI outputs during evaluation, and elicitation of stakeholder input for alignment. While such designs offer efficiency and scalability, they often undermine the very conditions under which disagreement can be epistemically fruitful---and worse, they risk entrenching perspectival homogenization. When individuals work without opportunities for collaboration, deliberation, or even minimal interaction,\footnote{As we discuss in the next subsection, such minimal interactions could include indirect forms of scaffolding, such as seeing dissenting justifications from others or beginning a task from a position already developed by a peer~\cite{page2017diversity}.} there is no chance for synergies to emerge from complementary perspectives, or for epistemically advantaged standpoints to inform collective understanding. Nor do such task designs allow individuals to face expectations that they need to provide reasons for their judgments or to engage in productive discourse. As a result, they preclude the benefits of both diversity and epistemically productive disagreement (see Sections~\ref{subsec:diversity},~\ref{subsec:standpoint}, and~\ref{subsec:disagreement}). Worse still, disagreements may be mischaracterized as noise rather than as useful first- or higher-order evidence (as in~\ref{subsec:preservation}). Take, for example, an evaluator who identifies a subtle privacy risk in AI output that others overlook, due to her knowledge of auxiliary datasets enabling composite attacks. The lack of interaction not only prevents others from learning from their peer; it also risks leading practitioners to interpret the divergent opinion as an idiosyncratic outlier rather than a crucial epistemic contribution. The use of majority voting to aggregate opinions compounds the issue by systematically excluding insights that only a minority of evaluators are positioned to provide.

Crucially, the epistemic benefits of disagreement cannot be recovered merely by increasing the diversity of task participants or refining aggregation procedures. Consider again recent work in participatory AI and pluralistic alignment. These approaches introduce novel frameworks and methodologies for incorporating a broader range of perspectives and for aggregating opinions in ways that---unlike simple majority voting---seek to avoid the tyranny of the majority~\cite{huang2024collective,openAI2024democraticAI}. However, many of these approaches still dissolve disagreement through consensus-seeking mechanisms that emphasize (group-aware) agreement in judgments---an orientation that is reflective of the ``bridging approach'' to deliberation that underpins them~\cite{small2021polis}. Because they do not afford participants with divergent perspectives meaningful opportunities to interact or engage in productive discourse, they risk missing the deliberative value of dissent.

A general upshot, then, is that realizing the epistemic benefits of disagreement requires taking seriously the \textit{sociotechnical} nature of AI development tasks, and, in particular, their \textit{organizational and social structure}.\footnote{The need for socially structured methods that respect disagreement is also emphasized in recent work on annotation practices~\cite[see][]{fleisig2024perspectivist}.} This means conceiving of task participants not as isolated individuals, but as embedded in networked communities of inquiry. It also requires seeing dissent not as a problem to be managed through aggregation, but as a potential catalyst for further epistemically productive processes. Recent efforts to enable interaction among annotators with divergent opinions~\cite{duan2020does}, to support communities in navigating divergent perspectives during dataset curation~\cite{kuo2024wikibench}, and to promote structured discussion during AI ruleset elicitation~\cite{bergman2024stela} all point toward promising directions. The framework proposed here offers a normative and methodological foundation for further advancing these efforts. Future works must also consider the critical role of social and organizational conditions under which communication takes place, attending in particular to the importance of democratic structure and egalitarian norms as \textit{enabling conditions} for the epistemic benefits of diversity and disagreement to arise~\cite{phillips2017real,longino1990science}. 

\subsection{How to navigate trade-offs: Modifying topology and communicating justifications}
\label{subsec:justification}
To be sure, efforts to harness the epistemic benefits of disagreement in AI development must confront a recurring challenge: fostering productive disagreement---especially in diverse collectives---can introduce inefficiencies, friction, or even conflict. Faced with these challenges, it might be tempting to design systems that suppress disagreement and prioritize agreement as a means of consensus-building. However, by attending to the specific mechanisms that underlie the epistemic benefits of disagreement discussed in Section~\ref{sec:why-disagree}, we can identify strategies that navigate these trade-offs. That is, rather than forgoing disagreement's epistemic benefits out of concern for its potential costs, we can design alternative strategies for effectively managing the trade-offs. 

In particular, we can think of two design levers: modifying the \textit{structure} of communication, and modifying the \textit{content} that is communicated. As discussed in Section~\ref{subsec:diversity}, research in network science has explored specific communication structures that can effectively balance the benefits of fostering communication and collaboration among diverse perspectives with the costs of communication and coordination~\cite{heydari2015efficient,heydari2020not}. These findings offer guidance for scalable task design in AI development in terms of workflows that structure \textit{how} communication occurs---for example, through network topologies that create targeted opportunities for exposure to disagreement where it is most epistemically valuable.\footnote{Note that the relevance of network structure applies to settings involving face-to-face discussion as well, insofar as such settings tend to instantiate a particularly topology, namely, a complete network in which everyone can in principle communicate with everyone else.}

A second approach involves modifying \textit{what} is communicated. Current AI design and evaluation practices typically solicit individual judgments---e.g., about which of two outputs is more appropriate, or whether and to what extent a post is toxic---seldom requesting the \textit{reasons} behind those judgments. As a result, the communicated content is limited to surface-level evaluations, without capturing the deeper reasoning processes that support them. Yet, as noted in Section~\ref{subsec:disagreement}, the epistemic value of disagreement often lies not in the judgments themselves, but in the justifications they elicit. Anticipated or active disagreement compels individuals to articulate reasons for \textit{why} they hold a view, revealing their underlying mental models and situated knowledge as well as the quality of their evidence. 

Consider a scenario where participants in annotation or model evaluation tasks are asked not only to assign a label, but to explain and justify their judgment, potentially in response to dissenting views. In such a case, the resulting justifications can provide important evidence. If some participants lack evidence or rely on fallacious reasoning, this information can lower the epistemic weight of their judgment. Conversely, if different individuals reach the same conclusion via independent but reliable reasoning paths or sources of evidence, the epistemic merit of that conclusion may be strengthened through considerations of methodological and evidential diversity (discussed in~\ref{subsec:preservation}).

While these epistemic benefits typically depend on the existence (or anticipation) of communication among disagreeing parties, they can still be realized by expanding the unit of communication in AI development tasks---from only individuals judgments to the justifications that support them. Recent efforts in machine learning to learn from additional human feedback, such as annotator rationales~\cite{zaidan2007using}, or to design platforms that support the elicitation and exchange of justifications~\cite{kuo2024policycraft} can help provide the infrastructure needed to make this shift. More broadly, future work can build on the proposed framework to examine how integrating justifications into the communication process---to produce richer training, evaluation, and alignment datasets that reflect contextually and causally relevant considerations---can impact the behavior of AI systems. 

\subsection{How to communicate disagreement}
The account of disagreement developed in this paper provides a normative grounding for recent work in the so-called perspectivist paradigm~\cite{fleisig2024perspectivist}. This line of research has sought to improve how diversity of perspectives and disagreement are handled in dataset construction (and the models trained on them) through methods such as: developing documentation artifacts that preserve information about annotator selection, compostion, and condition~\cite{diaz2022crowdworksheets}; tracking annotator-level labels~\cite{prabhakaran2021releasing}; introducing aggregation procedures and modeling techniques that defer resolution of disagreement until decision time~\cite{gordon2022jury,davani2021dealing,fleisig2023majority}; and leveraging measures of disagreement to provide users with additional higher-order evidence~\cite{davani2021dealing}.

Building on this valuable line of work, we suggest that future documentation practices should focus on ensuring \textit{coherence} across design decisions made at different stages of the AI development task. As the discussion above makes clear, the epistemic value of disagreement depends not on isolated design decisions, but on how those decisions relate to one another across the task. For instance, ensuring the presence of relevant perspectives in the antecedent stage (Section~\ref{subsec:why-perspectives}) is a prerequisite for realizing the benefits of disagreement in the process stage (Section~\ref{subsec:why-task}). Conversely, poorly designed task structures at the process stage lead to failures to leverage diverse perspectives and epistemically advantaged standpoints. Documenting these decisions in an integrated way, and assessing their coherence, is essential for providing meaningful transparency about how disagreement is dealt with in a task.

Attending to disagreement as higher-order evidence is also crucial for the communication and curation of diverse views in AI outputs. A growing body of research has begun to investigate how certain perspectives become underrepresented or obscured in the outputs of large language models~\cite{santurkar2023whose,stammbach2024aligning}. Relatedly, recent work raise concerns that in domains characterized by substantive disagreement AI systems exhibit pronounced sycophancy---the tendency of AI systems to echo user views~\cite{perez2022discovering,lindstrom2024ai}. That is, precisely where the communication of disagreement could help users calibrate their confidence or recognize contested issues, current AI systems often obscure dissent and reinforce an illusory consensus.

While recent work in pluralistic alignment has begun to address these concerns~\cite{sorensen2024roadmap}, the question of how to communicate epistemically meaningful disagreement in AI output remains normatively complex. The lines of work discussed here can offer helpful guidance. For example, drawing on work discussed in Section~\ref{subsec:diversity}, \citet{fazelpour2022diversityml} argue that how we operationalize diversity in different stages of the AI lifecycle---such as during output curation---should depend on prevailing contextual and normative considerations. These, in turn, inform which formal measures are most appropriate for implementation. For example, when competing viewpoints are equally justified, an egalitarian conception of diversity may be most apt, favoring a balanced presentation of views---formalized using entropy-based measures. In contrast, when there is justified consensus about an issue, a representative curation---formalized using proportionality measures---may be more appropriate, as balancing would unjustifiably amplify outlier views. Future work can build on this by designing mechanisms that present not only information about the existence of diverse views, but also about the type and variety of reasoning underlying them, as discussed in Section~\ref{subsec:preservation}. Such efforts would support disagreement's potential to serve as a form of higher-order evidence that can help both practitioners and users to more accurately assess the reliability of AI outputs. 
\section{Conclusion}
Taken together, the preceding sections offer a structured approach to realizing the epistemic benefits of disagreement in AI development. By organizing design decisions in AI development tasks, and linking them to relevant epistemic rationales for valuing disagreements, the framework provides practitioners with resources to reason about the epistemic benefits of disagreement in a principled and coherent way, and supports efforts to resist perspectival homogenization in AI design and governance processes. Beyond offering correctives to common, but mistaken assumptions, the framework provides normative grounding for a range of emerging methodologies, including works in participatory AI and pluralistic alignment. We hope that this work can spark future research on the design of sociotechnical infrastructure needed for better eliciting, preserving, structuring, and communicating disagreement, and realize its varied benefits.

\begin{acks}
The authors would like to thank the audiences at Stanford and Georgetown, and the anonymous reviewers at FAccT for their valuable feedback. SF was supported by Schmidt Sciences AI2050 Early Career Fellowship. 
\end{acks}

\bibliographystyle{ACM-Reference-Format}
\bibliography{refs}


\end{document}